# Undulatory Swimming in Sand: Subsurface Locomotion of the Sandfish Lizard

Ryan D. Maladen[1], Yang Ding[2], Chen Li[2], Daniel I. Goldman[1,2]

[1]Interdisciplinary Bioengineering Program, Georgia Institute of Technology, Atlanta, GA 30332, USA.
[2]School of Physics, Georgia Institute of Technology, Atlanta, GA 30332, USA.

Published in Science 325(5938), 314–318 (2009).
DOI: 10.1126/science.1172490

## Abstract

The desert-dwelling sandfish (Scincus scincus) moves within dry sand, a material that displays solid and fluidlike behavior. High-speed x-ray imaging shows that below the surface, the lizard no longer uses limbs for propulsion but generates thrust to overcome drag by propagating an undulatory traveling wave down the body. Although viscous hydrodynamics can predict swimming speed in fluids such as water, an equivalent theory for granular drag is not available. To predict sandfish swimming speed, we developed an empirical model by measuring granular drag force on a small cylinder oriented at different angles relative to the displacement direction and summing these forces over the animal movement profile. The agreement between model and experiment implies that the noninertial swimming occurs in a frictional fluid.

The locomotion of organisms (1, 2), whether by running, flying, swimming, or crawling, is the result of multiple-degree-of-freedom nervous and musculoskeletal systems interacting with an environment that often flows and deforms in response to movement. Nearly all experiments and models of terrestrial locomotion have been developed for running and walking on rigid, flat, no-slip frictional substrates for which the complication of substrate flow is not considered. In contrast, complexity in interaction with the environment in aquatic and aerial locomotion (swimming and flying) is well recognized (3). Determining mechanisms for propulsion or lift in these media is always possible in principle because the rules of interaction with fluids are worked out: They require solving Navier-Stokes hydrodynamics in the presence of moving boundary conditions.

A major challenge in biology is to understand the locomotion of organisms that walk, crawl, or burrow on or within terrestrial substrates like sand (4), soil (5), and muddy sediments (6) that display both solid and fluidlike behavior. In such materials, validated theories such as the Navier-Stokes equations for fluids do not exist, and visualization techniques [such as particle image velocimetry in fluids (7)] are nearly nonexistent. Understanding of the mechanics of subsurface movement has ecological importance and could reveal how the actions of small burrowing organisms can transform entire landscapes (8).

Animal burrowing and movement within granular media is relevant to desert organisms like scorpions, snakes, and lizards that move within sand to escape heat and predators and hunt for prey (9, 10). Desert sand [which covers 6 to 10% of land surface (11)] is an example of a granular material, a collection of dissipative particles that interact through contact forces and in bulk can display solid and fluidlike features (12) when disturbed. A key parameter that controls the response of granular media to intrusion is the volume fraction $\phi$, the ratio of material volume to total occupied volume. In dry granular media in natural environments, $\phi$ depends on the history of the sand (for example, perturbations by wind or animal burying and digging), and can vary between 0.57 and 0.64 (13). The response of granular media to intrusion depends on $\phi$: Closely packed material at high $\phi$ must expand to flow, whereas loosely packed material at low $\phi$ consolidates (14). The effects of $\phi$ on drag are largely unexplored, although we have recently found that vertical penetration resistance doubles as $\phi$ increases by just 0.08 (15).

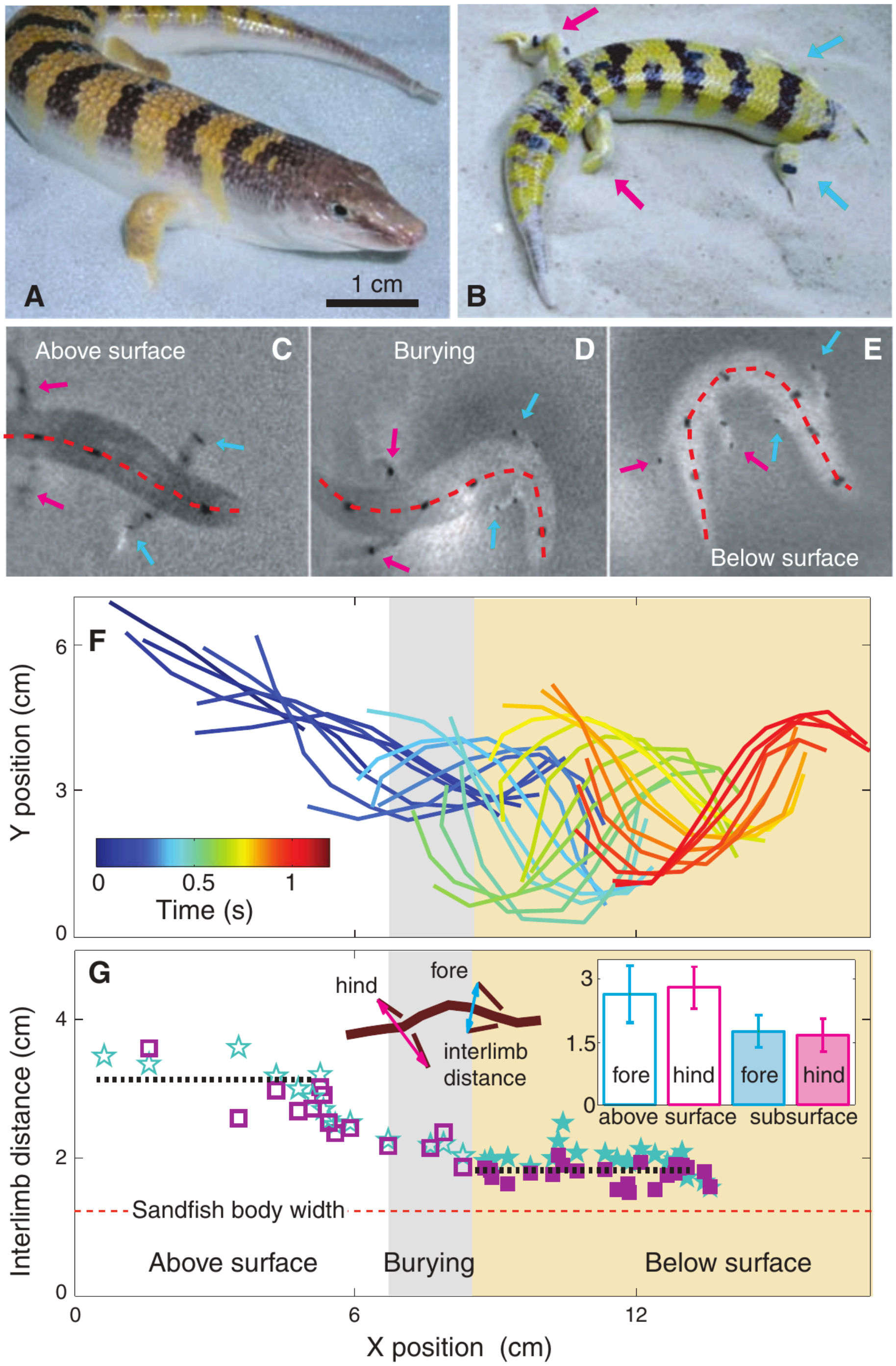

**Figure 1.** Burial and undulatory subsurface swimming of the (A) sandfish lizard. (B) The burying sandfish with opaque makers bonded to its fore (cyan) and hind (purple) limbs and midline, just before it dives into the media. (C) Representative x-ray images of the sandfish as it moves with a diagonal gait on the surface, (D) as it buries into the media with high-amplitude oscillations and use of limbs, and (E) moving subsurface with its limbs at its sides. (F) Curves show the tracked midline of the sandfish as the animal runs on the surface using a diagonal gait (1, 9) with little back bending (unshaded region), buries with significant back bending (gray region from when snout first enters material to when fully submerged), and then swims within the media with large-amplitude back bending (brown shading). The curve color (color bar, inset) indicates time progress of the motion. (G) Interlimb distance of the fore (blue star) and hind (purple square) limbs versus the position of the center of mass. The dashed lines represent the average interlimb distance (10 animals) above and below the surface. (Inset) Bar graph shows the interlimb distance (mean ± SD) for the fore and hind limbs both above and below the surface. n = 46 trials were recorded.

To investigate how rheological features of the material influence the locomotor mode and performance of an organism moving within sand, we used high-speed x-ray imaging to study a small (~10 cm) desert-dwelling lizard, the sandfish, that inhabits the Saharan desert of Africa and moves within granular media of different $\phi$. The sandfish's above-ground burial process has been described (9), and it is hypothesized that its counter-sunk lower jaw and smooth scales with low friction and low wear properties (16) aid swimming and digging. However, little is known about how the animal moves subsurface. Although it has been hypothesized that body motion plays an important role in thrust production (9, 17) in sand-dwelling lizards, a recent study using nuclear magnetic resonance (NMR) to visualize subsurface motion proposed that the sandfish used its limbs in a paddling motion along with undulations on its body to generate thrust subsurface (18). However, the observational quality and number of trials were limited. We show through detailed biological experiment and development of a model of the locomotion that subsurface sandfish motion is not driven by limbs but by an undulatory gait without use of limbs. Thrust and drag result from frictional grain flow, which leads to features of the swimming that are intermediate to movement in fluids and on solid surfaces, and surprisingly are independent of $\phi$.

To perform controlled studies of the burial and swimming of the sandfish (11 animals; mass = 16.2 $\pm$ 4 g, body length $L$ = 8.3 $\pm$ 3.3 cm, defined as the length from snout tip to the base of the tail), we used a fluidized bed (fig. S1) (19) to prepare a 10-cm-deep container of 0.27 $\pm$ 0.04 mm glass beads into naturally occurring loosely packed and closely packed volume fractions $\phi$ = 0.58 and $\phi$ = 0.62, respectively. The sandfish took 672 $\pm$ 239 ms to complete its entire burial process, independent of volume fraction (Student's $t$ test, $P > 0.05$) (movie S1). Once below the surface, the dorsal x-ray video imaging (movies S2 and S3) revealed that the animal no longer used limbs for propulsion. Instead, the animal placed its limbs against its sides and executed an undulatory motion (Fig. 1, E and F) with large-amplitude axial oscillation, using its body to propel itself at speeds of up to ~10 cm/s. We observed similar limbless undulatory motion in beads with mean diameters ranging from 0.1 mm to 3 mm. To quantify the lack of limb use, we measured the distance between pairs of limbs (fore and hind) (Fig. 1G). As the animal ran across the surface, this interlimb distance oscillated around 2.75 $\pm$ 0.59 cm, indicating the sprawl of the limbs during the diagonal gait (1, 9). As it began to bury, the interlimb distance decreased significantly (Student's $t$ test, $P < 0.01$) until it became a constant (1.72 $\pm$ 0.53 cm) close to body width (1.23 $\pm$ 0.51 cm) because limbs were placed close to the side of the body. Like in (18), side-view x-ray imaging of the sandfish subsurface showed that the animal undulates in a plane at ~ 22.2 $\pm$ 3.7° with respect to the horizontal and buries consistently to a depth of 2.1 $\pm$ 0.5 cm (surface to top of mid-point of the back), independent of volume fraction (Student's $t$ test, $P > 0.05$).

We characterized the undulatory kinematics by digitizing the midline from the dorsal x-ray images (Figs. 1F and 2A). The undulatory motion fit well ($r^2 > 0.85$, $P < 0.01$) the form of a posteriorly traveling single-period sinusoidal wave (Fig. 2A) such that $y = A \sin 2\pi(x + v_w t)/\lambda$, with $y$ the displacement away from the midline of a straight animal, $A$ the amplitude, $\lambda$ the wavelength, $f$ the wave frequency, and $v_w = f\lambda$, the wave speed. From the fit, we deduced these parameters as well as the forward swimming speed of the center of mass $v_x$ (which was not equal to the wave speed; see discussion below). During a swim, there was no change in either $A$ ($P = 0.12$) or $\lambda$ ($P = 0.66$), and neither $A$ nor $\lambda$ varied significantly for the different $\phi$ ($P > 0.05$ comparing each treatment by using a repeated measures one-way analysis of variance) (Fig. 2B). The large-amplitude undulations over the entire body are unlike the kinematics of other undulatory swimming organisms on the sandfish-sized scale [such as eels (20)], which tend to increase in amplitude posteriorly. The undulations do resemble those of eel movement on land (20) as well as small (<1 mm) swimmers (for example, nematodes) in fluids at low Reynolds (*Re*) number, in which the inertia of the material surrounding the organism plays no role (21, 22).

The sandfish increased swimming speed by increasing wave frequency $f$ (Fig. 2C) (Student's $t$ test, $P < 0.01$), and the slope of the $v_x$ versus $f$ relationship was independent of volume fraction (Student's $t$ test, $P > 0.05$). We found, however, that $v_x$ is not simply the product of $f$ and $\lambda$, but instead the slope is reduced by a constant factor $\eta$ so that $v_x = \eta f$ (Fig. 2C) with $\eta$ = 0.54 $\pm$ 0.12 and $\eta$ = 0.49 $\pm$ 0.09 for loosely and closely packed material, respectively. Because $v_w = \lambda f$, this implies that the average forward speed is a fraction of the wave speed. Thus, the animal is not moving in a tube because successive undulations do not trace a continuous path; this can be seen in Fig. 1F. Instead, tracer particles placed in the bulk reveal that there is backward displacement of material as the animal moves forward. Slipping while progressing is common to undulatory swimmers in deformable media across length scales (for example, eels and spermatozoa in fluids) (1) and is characterized by the wave efficiency (η) (21), defined as the ratio between the forward speed of the animal ($v_x$) and the velocity of the wave traveling down its body ($v_w$): $\eta = v_x / v_w$. Because λ for the sandfish is independent of $f$ in our experiments, η is the slope of the $v_x/\lambda$ versus $f$ curve and is shown in Fig. 2C. Typical wave efficiencies of organisms in moving in fluids at low $Re$ (such as nematodes) are 0.25 (23), nearly a factor of two smaller than the sandfish, whereas organisms undulating (creeping) along the solid surface of an agar-air interface can have η close to unity [$\eta \sim 0.8 - 0.9$ (21)] because deformations in the surface allow movement that is effectively in a tube. η for the sandfish in granular media is intermediate to that for fluids and solids and is independent of ϕ, even though materials with different ϕ have penetration resistances that can vary by a factor of 2 (15).

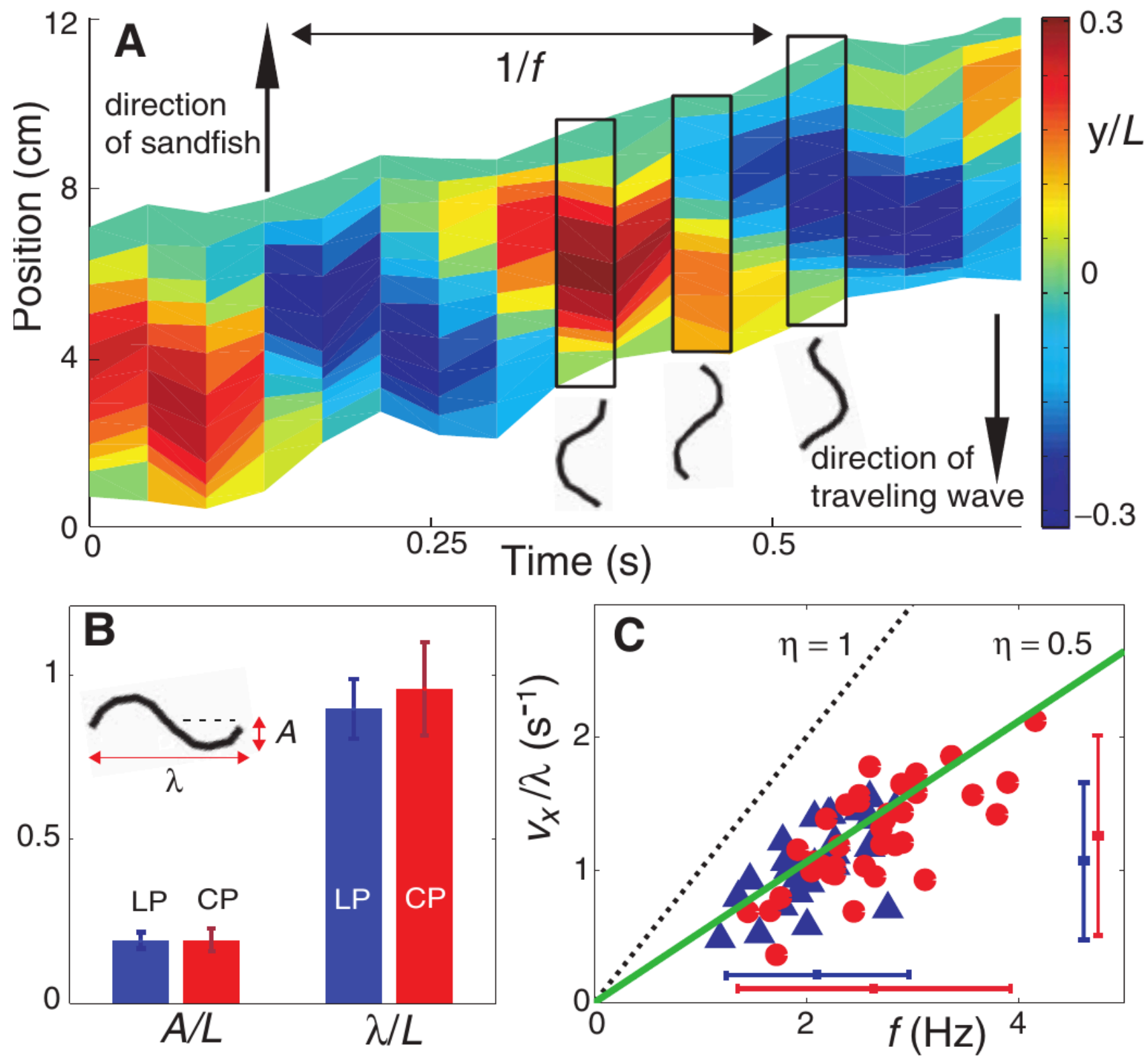

**Figure 2.** Kinematics of the undulatory sandfish motion. (A) Traveling wave moving down the body of the sandfish opposite to the direction of the sandfish forward motion (sampled every 0.04 s). For each time instant, the instantaneous lateral displacement of a tracked section of the sandfish is represented in color. The black curves represent the tracked midline (for example, Figure 1E, snout tip to tail tip) of the sandfish. (B) The amplitude A and wavelength λ of the sinusoidal fit to the tracked midline of the sandfish (inset), for loosely and (C) Red circles (closely packed, n = 24 trials), and blue triangles (loosely packed, n = 22 trials) show dependence of swimming velocity on wave frequency. The slope of green line is the approximate wave efficiency $\eta \approx 0.5$ of animal (Fig. 4A) and the dashed black line represents η = 1. The horizontal and vertical blue and red bars (mean ± 2 SD) represent the range of frequencies and forward velocities (normalized by λ) accessed in loosely packed (2.10 ± 0.75Hz, 0.98 ± 0.27 per second) and closely packed (2.63 ± 1.13 Hz, 1.28 ± 0.37 per second), respectively. The average frequency and average speed are greater in closely packed than in loosely packed material (Student's t test, $P < 0.01$). n = 46 trials were recorded.

Prediction of η for undulatory movement in fluids and solids has been accomplished by use of models of interaction of the organism and the environment: In low $Re$ Newtonian fluids, forces are purely viscous (Stoke's law) such that force is proportional to instantaneous velocity (22, 23), and in undulatory crawling on solids force is determined with static/dynamic friction coefficients (22). Although the equivalent force laws are not available in granular media, features of the sandfish motion allow us to hypothesize about the medium in which it swims: (i) It is well known that for slow enough disturbances, the drag in granular systems is dominated by friction between flowing grains, and thus this makes drag independent of velocity (14, 24–26). (ii) At the observed body speeds and accelerations during undulatory motion, inertial forces due to accelerating material are small as compared with frictional drag forces ($ma/F \approx 0.2$, where $F \approx 2$ N is estimated for a small cylinder dragged in sand and $ma \approx$ 0.4 N is found from the product of the acceleration of the sandfish body and a mass of sand surrounding it), and thus inertia can be neglected. (iii) Because grain interactions are dissipative, disturbances typically dampen in short times, and thus changes in the static configuration dominate the drag force.

Therefore, we hypothesize that the animal swims in a so-called "frictional fluid" (27), such that flowing grain-grain and grain-animal friction determine the thrust and drag forces. Such a frictional fluid superficially resembles low $Re$ swimming in that there is no inertia, but the mechanism for drag is frictional, rather than viscous. Movement in a frictional fluid differs from noninertial crawling on a surface in which anisotropic friction [originating from, for example, deforming the surface and pushing off as does nematode on agar (21) or using overlapping belly scales or changes in weight distribution as does a snake (28)] generates net propulsive force without net flow of material.

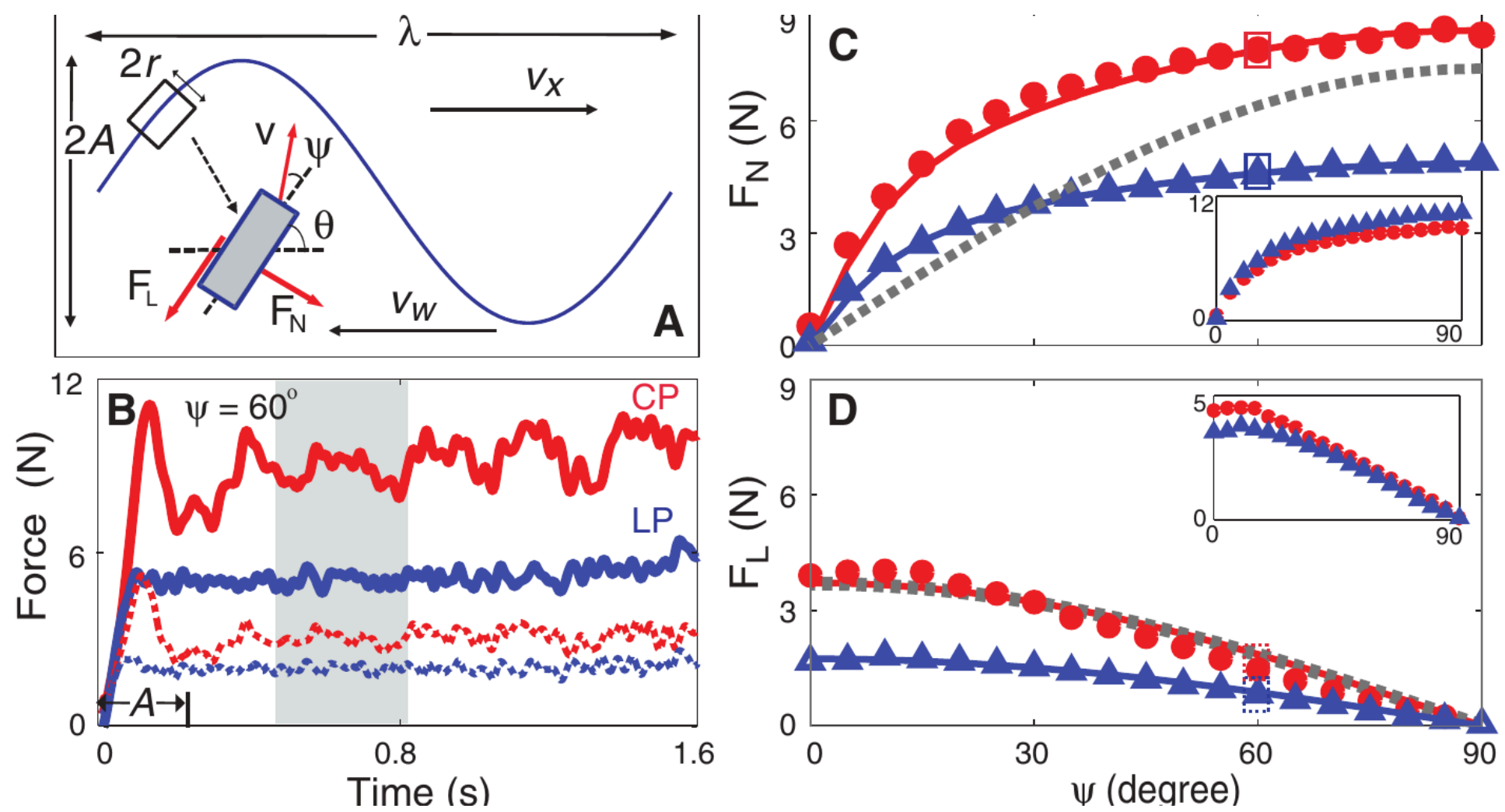

**Figure 3.** Drag measurements to model undulatory locomotion: (A) Diagram of forces F (perpen- N dicular) and F (parallel) on a given element of a sinusoidal traveling wave. (B) Representative total L drag force profiles (including sidewalls and end-faces) versus time for a stainless-steel cylinder (diameter = 1.58 cm and length = 4.00 cm) oriented at an angle ψ = 60° to the direction of its forward velocity (10 cm/s). The solid lines correspond to F and the dashed to F . The red and blue N L colors correspond to closely and loosely packed media, respectively. The label A represents displacement associated with the average amplitude of oscillation of the sandfish. (C) Average F N and (D) F on the cylinder (removing force contribution from the end-faces); red circles and blue L triangles correspond to closely and loosely packed preparations, respectively. In (C) and (D), the average forces corresponding to the force profiles in (B) are indicated with the respective rectangles. Both insets show the loosely packed force data scaled by 2.1. Averages of F and F are N L calculated over the shaded region in (B). Solid lines are model fits described by Eq. 1. For comparison, dashed gray lines correspond to F and F calculated for a fluid (water) by choosing a N L velocity that fits F versus ψ.

To test this hypothesis, we developed a model inspired by similar local force, noninertial models [typically called Resistive Force Theory (RFT) (23)] such that the body of the organism is partitioned into elements, each of which generates thrust and experiences drag when moving through a media and whose forces are not influenced by force fields from other segments. If $F_N$ and $F_L$ are the normal and lateral forces (perpendicular and parallel to the element respectively) acting on the sides of the sandfish body, net forward force on the element can be written as $F_x = F_N \sin\theta - F_L \cos\theta$, where θ is the angle between the forward direction of the animal and the orientation of an element of the organism (Fig. 3A). The forces for each element are integrated over the length of the body (snout to tail tip) and for a full period of a traveling wave to obtain average swimming speed. Unlike in fluids, we have no validated theory to calculate the force on an intruder as a function of its angle relative to displacement or for varying $\phi$. Therefore, we used a combination of measurements (19) of the granular thrust and drag forces on a stainless steel cylindrical rod [with grain-surface friction coefficient similar to the sandfish (16)], insights from Fig. 3B, and an empirical fitting function that incorporates drag on the sidewalls and end-faces of the cylinder (equation S8) so as to approximate $F_N$ and $F_L$ for elements of the sandfish body as:

$$F_N = 2lr(C_S \sin\beta_0 + C_F \sin\psi)$$
$$F_L = 2lrC_F \cos\psi \qquad (1)$$

where $\tan\beta_0 = \cot\gamma_0 \sin\psi$, $\psi$ is the angle of the segment with respect to its local velocity vector, $l$ and $r$ are the length and radius of the segment (assumed cylindrical), and the constants $C_S$, $C_F$, and $\gamma_0$ characterize the material response to drag. Plots of $F_N$ and $F_L$ as a function of $\psi$ are shown in Fig. 3, C and D. Equation 1 represents the data well ($R^2 = 0.97$) (solid lines shown in Fig. 3, C and D), and the fit coefficients are given in table S1.

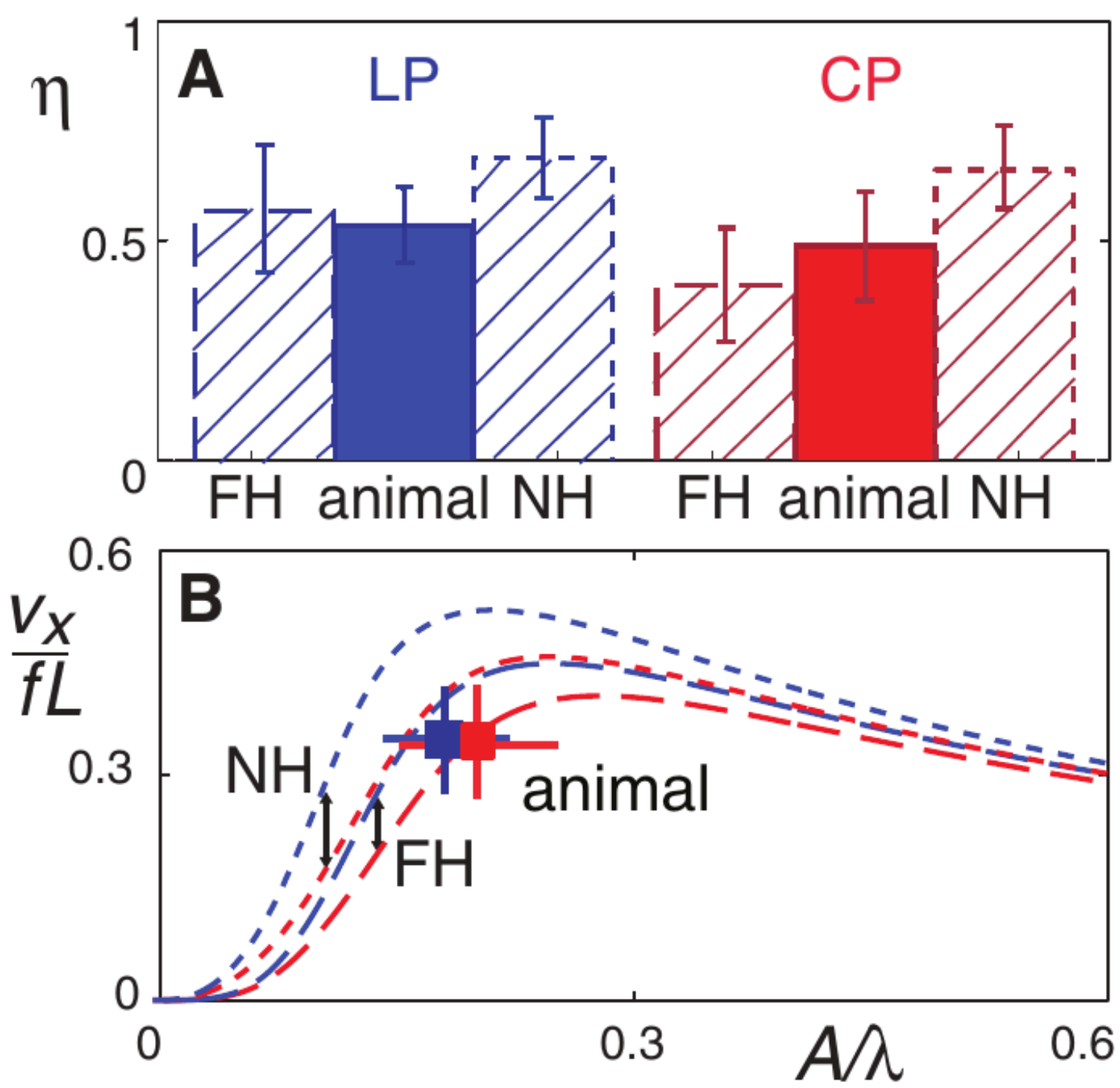

**Figure 4.** Comparison of experimental results and model predictions; blue denotes loosely packed and red denotes closely packed material. (A) Animal wave efficiencies from experiment are represented by solid bars (loosely packed average $\eta = 0.54 \pm 0.12$; closely packed average $\eta = 0.49 \pm 0.09$). Shaded bars correspond to model predictions assuming head drag of a flat disk (FH) and no head drag (NH). (B) Predicted swimming speed (in bodylength per cycle) versus amplitude normalized by wavelength. The dashed lines are model predictions with no-head (short-dashes) and flat-head (longdashes) for loosely packed and closely packed material. Squares denote the measured average animal swimming speed for loosely packed and closely packed material. The range of frequency and speed accessed is given by the extent of the solid lines. n = 46 trials were recorded.

$F_L$ is well approximated by the cosine function, similar to viscous fluids or frictional material, indicating that its source is the resolved component of friction between the grains and the body. The form of $F_N$ is not well approximated by the corresponding sine component (Fig. 3C, dashed gray lines) but rather increases more rapidly for $\psi < 30°$ and increases more slowly for $\psi > 30°$. We hypothesize that above this angle [close to the angle at which internal slip planes or shear bands form and move (14)], the force increases slowly because of the formation of a solid region of grains that moves along with the rod (14, 26). Preliminary numerical simulation indicates that the size of the solid region grows slowly with ψ for ψ > 20°. The fitting function (Eq. 1) incorporates these effects in $\bar{F}_N$ and $\bar{F}_L$ through terms proportional to the angular component of $C_S$ (which is determined primarily by grain-object frictional forces) and an extra term in $\bar{F}_N$ that increases rapidly and then saturates for ψ > 20°. In this term, the coefficient $C_S$ is set by grain-grain frictional force and $\gamma_0$, which is related to the internal slip angle. $C_F \approx C_S/2$, indicating that grain-grain friction is larger than grain-surface friction.

To determine the wave efficiency (and thus swimming speed for a given frequency), we balanced thrust and drag over a cycle (net force, $\overline{F}_x = 0$ in equation S7, assuming constant cycle average speed $v_x$). Because $v_x$ and $v_w$ appear only as a ratio (from the kinematics in equation S2) in the argument of the integral in Equation S7, and do not appear in the $\bar{F}_N$ and $\bar{F}_L$ terms (fig. S2B), the force balance predicts η. Numerical integration of the model over a measured organism sinusoid from Fig. 2 with constant cylindrical crosssection predicts 0.4 ≤ η ≤ 0.65 independent of ϕ (Fig. 4A), which is in accord with experimental observation; the bounds on η are set by assuming the head drag is zero or that of a flat plate because measurements for the drag force on an object with the exact morphology of the shovelshaped head are unavailable. The model predicts a maximum in forward progress per cycle when $A/\lambda \approx 0.2$, and we find that the measured kinematics fall close to this speed (Fig. 4B). The maximum is due to a competition between increased η (less slipping) from increased $A$ and decreased λ because for an inextensible animal λ decreases as $A$ increases. An analytic lowamplitude approximation (equations S9 to S14) demonstrates the relevant scaling of η.

It is remarkable that η does not change significantly for different ϕ, especially because drag forces between closely and loosely packed material differ by ≈ 80%. The model indicates that η is essentially constant because $\bar{F}_N$ and $\bar{F}_L$ scale by the same ratio of 2.1 between closely and loosely packed material (Fig. 3, C and D, insets), and it is this ratio that determines η (equation S14). The model implies that η is greater than in low-$Re$ fluids because of the greater thrust to drag force ratio in the frictional granular media.

The agreement between experiment and model indicates that the assumptions of velocity independence, local interaction, and the use of the average $\bar{F}_N$ and $\bar{F}_L$ for all phases of the motion are good approximations at these slow swimming speeds. Although the assumption that all body and head segments encounter material with the same ϕ is probably not true (because the body encounters grains disturbed by the head and other body segments), because $\bar{F}_N$ and $\bar{F}_L$ scale in the same ratio in all ϕ and this ratio determines η, wave efficiency should be independent of position along the body. The model assumptions break down in the approximation that the instantaneous force (Fig. 3B) can be replaced by the average force. This is a good approximation except for $F_N$ in closely packed material (Fig. 3B, red curves); enhancement of $F_N$ because of peak stress significantly greater than the mean could explain why, contrary to expectation, the range of $f$ (and $v_x$) that the animal accessed was greater in closely packed material and the maximal swimming speed the animal was able to access was ≈50% larger in closely packed material (Fig. 2C).

In conclusion, we used high-speed x-ray imaging to show that the sandfish lizard propels itself within granular media using a largeamplitude traveling-wave oscillation of its body without using its limbs and that the wave efficiency of this motion is independent of the volume fraction of the medium. Using a model that balances thrust and drag with measured granular drag laws, we can predict the wave efficiency and optimal kinematics. We demonstrated that the animal swims in a medium with properties intermediate to those of low $Re$ viscous fluids and frictional solid surfaces and that are unique to dense granular beds, namely velocity-independent forces and enhanced normal-drag forces. Our results demonstrate that burrowing and swimming in complex media can have similar complexity to movement in air or water, and that organisms can exploit the solid and fluidlike properties of these media to move effectively within them.

29. This work was supported by NSF Physics of Living Systems grant PHY-0749991 and the Burroughs Wellcome Fund. We thank R. Full and T. Papenfuss for helpful discussion. We thank S. Steinmetz and N. Gravish for experimental assistance and P. Umbanhowar for discussion of the drag model and careful reading of the manuscript.

## Supporting movies

1. Movie S1: Visible-light recording of burial
2. Movie S2: X-ray recording of subsurface swimming
3. Movie S3: X-ray recording with body and limb markers

## Supporting Material

Corrected supporting material, 5 November 2009: missing square roots in Equations S4, S7, and S12 restored; Equation S11 labeled; analytic-solution section title corrected.

## Materials and Methods

## Sandfish X-ray Experiments

Sandfish lizards (Scincus scincus) were obtained from a commercial supplier (LLL Reptile, CA, USA), housed in a dedicated animal care facility, and kept in a 12 hour day/night light cycle. The average snout vent length (SVL), which we denote as body-length $L$ in the manuscript and which is $\approx$ 75% of the snout-tip to tail-tip length, was 8.3 $\pm$ 3.3 cm. The average mass was 16.2 $\pm$ 4 grams. High speed (250 fps) x-ray and visible light video of the sandfish were taken as they moved on and within a 10 cm deep layer of granular material (0.27 $\pm$ 0.04 mm diameter spherical glass beads, density $\rho \approx$ 2.4 $g/cm^3$ ) held in a container (21.5 cm $\times$ 18 cm). The granular material was prepared in two volume fractions ($\phi$ = 0.58 and $\phi$ = 0.62), loose and close packings for dry grains. The material state was set using an air fluidized bed, a counter flow of air to a collection of granular media. Below a critical flow rate (the onset of fluidization), grains are stationary; above it, the material is fluidized and grains flow. To generate loosely packed states ($\phi$ = 0.58) the flow is increased to above the fluidization onset and then slowly decreased to zero. To generate closely packed states the air flow is repeatedly pulsed. Increasing the number of pulses increases $\phi$; typically we delivered over 200 pulses to achieve the closely packed state ($\phi$ = 0.62).

The x-ray system (Radiological Imaging Systems, OEC 9000) energy was varied in the range of 85 $-$ 100 kV at 20 mA. After the x-ray beam was turned on, a gate to a holding pen was rapidly raised to release the animal onto the surface of the granular media. A thin aboveground plexiglas barrier (visible in Movies S1 and S2) forced the animal into the dive at the same location in the bed. The animal consistently took a few steps and then dove into the material. The sandfish appears dark in the above ground x-ray images since the radiation must travel through the sand and the animal, while below ground it appears lighter since sand has been displaced horizontally by the body of the organism. Experiments were conducted between 30 and 35 °C.

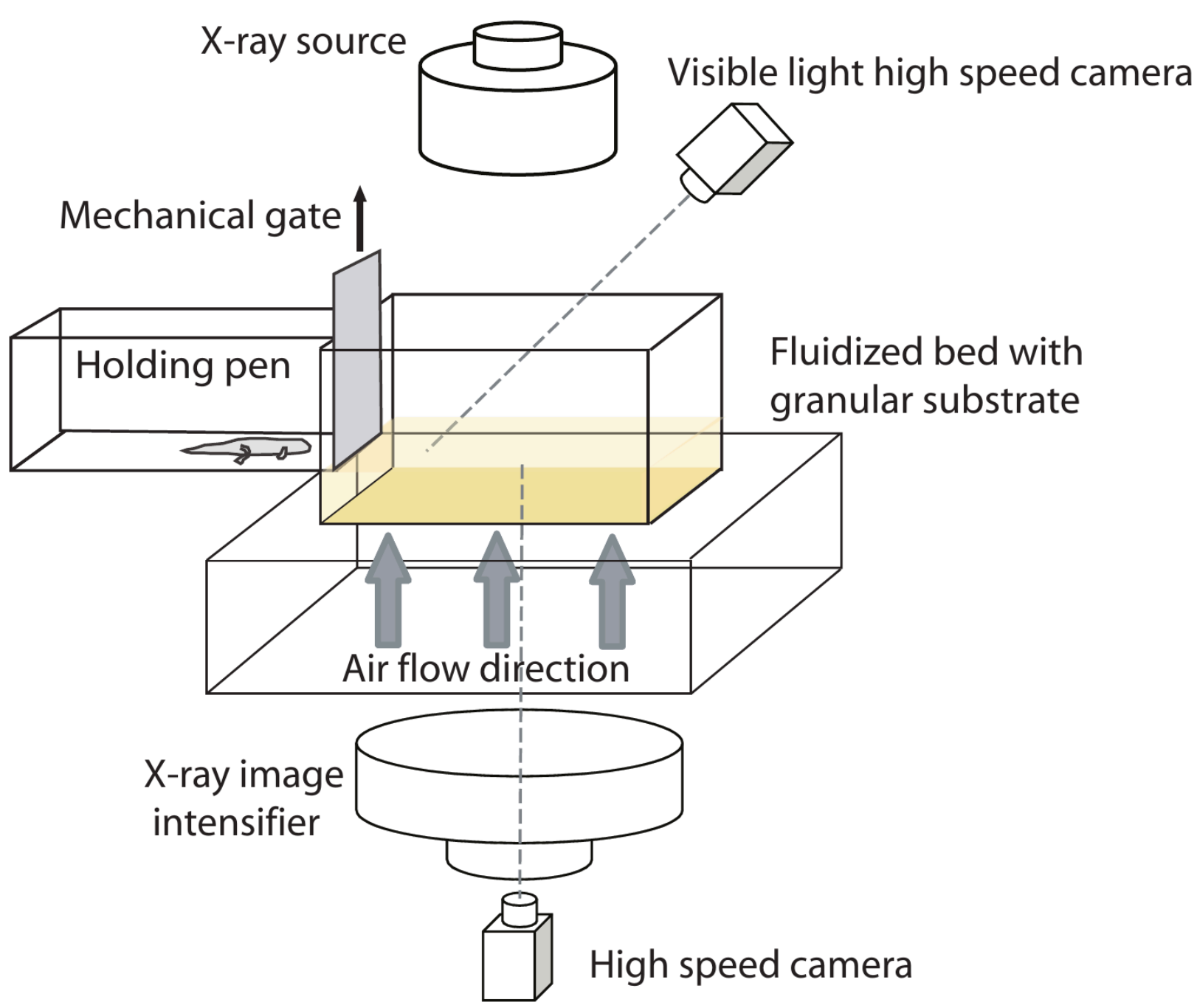

**Figure S1.** Experimental setup: a fluidized bed is placed between the source and image intensifier of a c-arm x-ray system. The initial state of the granular media (0.3 mm spherical glass beads) used to fill the fluidized bed is set using air pulses. Once the material is prepared the sandfish is released from the holding pen using a mechanical gate. Simultaneous x-ray and visible light video (250 fps) are recorded as the animal moves above and then below the surface. Drawings are not to scale.

The x-ray (Movies S2 and S3) and visible light (Movie S1) videos were synchronized to determine when the sandfish was moving subsurface. The temporal frequency and forward velocity of the animal were measured from this section of video. Displacement was calibrated using a grid. X-ray visible natural body markers (the snout and lungs) as well as the silhouette of the animal were used to estimate the mid-line of the animal and track it using MATLAB. The digitized undulation profiles of the animal were fit with the equation of a sinusoidal traveling wave to obtain the spatial and temporal characteristics of the animal. We collected representative runs with opaque markers bonded to the body midline and limbs of the sandfish to verify the lack of limb use subsurface (Movie S3). The total mass of the markers was 0.04 g, much less than the 16 g animals, The inter-limb distance measure allowed us to determine the nature of limb use subsurface and contrast it with the diagonal gait used in surface locomotion. The visible light video was used to determine the sandfish burial time.

Our model (explained later) suggests (Fig. S3) that the variability in the animal (the spread in velocity for a given frequency) is a result of the sensitivity of the swimming speed to small changes (within experimental variation during a swim) in both the amplitude and the wavelength.

## Resistive Force Theory for Granular Media

We developed our Resistive Force Theory for sandfish locomotion in granular media based on work by Gray & Hancock (S1) for non-inertial low Re swimmers. The animal is modeled as a flexible cylinder with a sinusoidal traveling wave progressing from head to tail (Fig. 3A) such that:

$$y = A \sin \frac{2\pi}{\lambda}(x + v_w t), \tag{S1}$$

where $y$ is the displacement away from the midline of a straight animal, $A$ the amplitude, $\lambda$ the wavelength, $f$ the wave frequency, and $v_w = f\,\lambda$ the wave speed. For a given forward velocity $v_x$, other quantities characterizing the motion can be calculated:

$$\begin{aligned} v_y &= \frac{dy}{dt} = \frac{2A\pi v_w}{\lambda} \cos \frac{2\pi}{\lambda}(x + v_w t) \\ \tan\theta &= \frac{dy}{dx} = \frac{2A\pi}{\lambda} \cos \frac{2\pi}{\lambda}(x + v_w t) \\ \psi &= \arctan(\frac{v_y}{v_x}) - \theta, \end{aligned} \tag{S2}$$

where $\theta$ is the angle of the axis of an infinitesimal cylindrical element with respect to the forward direction and $\psi$ the angle between the axis of the element and its instantaneous velocity.

We assume that the forces $F_N$ and $F_L$ (see Fig. 3A for the force diagram, and text and Rod Drag Empirical Fitting Function section for discussion of the force model) on an infinitesimal element of the animal are proportional to the area of the longitudinal cross section $\delta A = 2r\delta s$ and are functions of $\psi$. The resultant force on the element in the forward direction (+$x$-axis) is:

$$\delta F_x = (P_N \sin\theta - P_L \cos\theta) 2r\delta s, \tag{S3}$$

where $P_N$ and $P_L$ are the forces per cross section area normal and parallel to the axis of the element. Then the total force on the body $F_{bx}$ in the direction of motion can be expressed as

$$F_{bx} = \int_0^\lambda (P_N \sin\theta - P_L \cos\theta)\sqrt{1+\tan^2\theta}2r dx. \tag{S4}$$

Although the angle of each element changes in time, since the integral is over one full wavelength the integral in Equation S4 is time invariant.

Equation S4 accounts for forces acting on the body but not the head. To account for the head we use two limits, a flat head that produces maximum drag and no head which produces no drag. For the flat head model (FH), we assume that the drag on the head $F_h$ is the same as the drag on the flat end face of a rod in our experiment. This drag is decomposed into two orthogonal components $F_N^T$ and $F_L^T$ and its form is approximated by Equation S8, setting the length $l$ = 0 and $r$ as the average radius of the sandfish. The net force in the forward $(+x)$

$$F_{hx}(t) = F_N^T \sin\theta - F_L^T \cos\theta. \tag{S5}$$

The drag force on the head is set to zero for the no head model (NH).

The time average of the force on the head ($\bar{F}_{hx}$) is equal to the spatial average:

$$\bar{F}_{hx} = \frac{1}{T}\int_0^T F_{hx} dt = \frac{1}{\lambda}\int_0^\lambda (F_N^T \sin\theta - F_L^T \cos\theta)dx. \tag{S6}$$

Combining the head drag with the body force we obtain the total time averaged force on the

$$\begin{aligned} \bar{F}_x = \int_0^\lambda & [(P_N(\psi)\sin\theta - P_L(\psi)\cos\theta)\sqrt{1+\tan^2\theta}2r \\ & + (F_N^T \sin\theta - F_L^T \cos\theta)/\lambda]dx. \end{aligned} \tag{S7}$$

At constant average velocity, $\bar{F}_x$ is zero. The forward velocity $v_x$ can then be found numerically if $v_w$, $A$, $\lambda$, $P_N(\psi)$ and $P_L(\psi)$ are known. Since the wave and body velocities in Equation S7 only appear as ratios (see Equation S2) and do not appear in expressions for $P_N$ or $P_L$, a proportional relation between $v_x$ and $v_w$ is expected. We can thus solve for the wave efficiency $\eta = v_x/v_w$. Figure S3 shows that $\eta$ increases rapidly with increasing $A/\lambda$ for $A/\lambda \lessapprox$ 0.3 and increases slowly for $A/\lambda \gtrapprox$ 0.3.

## Drag Experiments in Granular Media

To obtain expressions for the forces on the sandfish, we moved a stainless steel cylindrical cylinder representing a sandfish body element through granular media at constant velocity (10 cm/s) and measured the resulting normal and lateral forces for angles $\psi$ between the element and its displacement direction ranging from $0°$ to $90°$ (Fig. 3 C and D, and Fig. S2 C). We used a fluidized bed (filled to a depth of 12 cm) to set the initial volume fraction ($\phi$) of the media. We chose stainless steel because a previous study (S2) as well as our measurements have shown that sandfish skin and stainless steel have about the same friction coefficient with sand ($\sim$ 0.2). The diameter of the cylinder ($2r$ = 1.58 cm) is similar to that of the sandfish body (1.75 cm). To avoid effects on the drag due to the finite size of the container, the cylinder is small compared to the size of the fluidized bed but large compared to the size of the grains. The cylinder is attached to a robotic arm (CRS Robotics) via a thin but stiff supporting rod which moves the structure at constant depth and velocity while a 6 d.o.f. force sensor (ATI industrial) mounted between the robotic arm and the supporting rod measures the net force generated during the drag motion (accurate to 0.06 N). The force on the supporting pole is also measured separately and subtracted to obtain the force on just the cylinder.

Previous studies show that drag in granular media is independent of speed (S3, S4) and increases proportionally to depth (S3). To test if this effect persisted in the regime relevant to the sandfish (0-40 cm/s), we dragged the cylinder perpendicular to velocity at 1, 5, 10, 20, and 30 cm/s (which covers the sandfish's range of speeds) at a fixed depth of 7.62 cm and found that over an order of magnitude change in speed, force only changed by 10% (Fig. S2 B); consequently we assumed force is independent of speed. We also dragged the cylinder oriented perpendicular to velocity at different depths (1.27, 2.54, 3.81, 5.08, 6.35, and 7.62 cm) at 10 cm/s and confirmed that as seen previously (S3) drag is proportional to depth (Fig. S2 A). Therefore, we measured drag force on the cylinder as a function of angle (for input into the Resistive Force Theory model) by dragging the cylinder at a fixed depth (7.62 cm) and a constant speed of 10 cm/s for different $\phi$.

Before each trial, the initial state of the medium was set using a fluidized bed and the test rod rotated to the desired angle using the robotic arm. For each angle the rod was pushed into the material to a depth of 7.62 cm and, after a 2 second pause, was dragged for 15.24 cm through the medium and then extracted. Both the force and position of the rod were recorded at 140 Hz. The forces parallel $(F_L)$ and perpendicular $(F_N)$ to the surface of the rod were resolved by the force sensor and averaged over the steady state drag region (Fig. 3 B). Three trials were conducted for each $\psi$ and $\phi$.

## Rod Drag Empirical Fitting Function

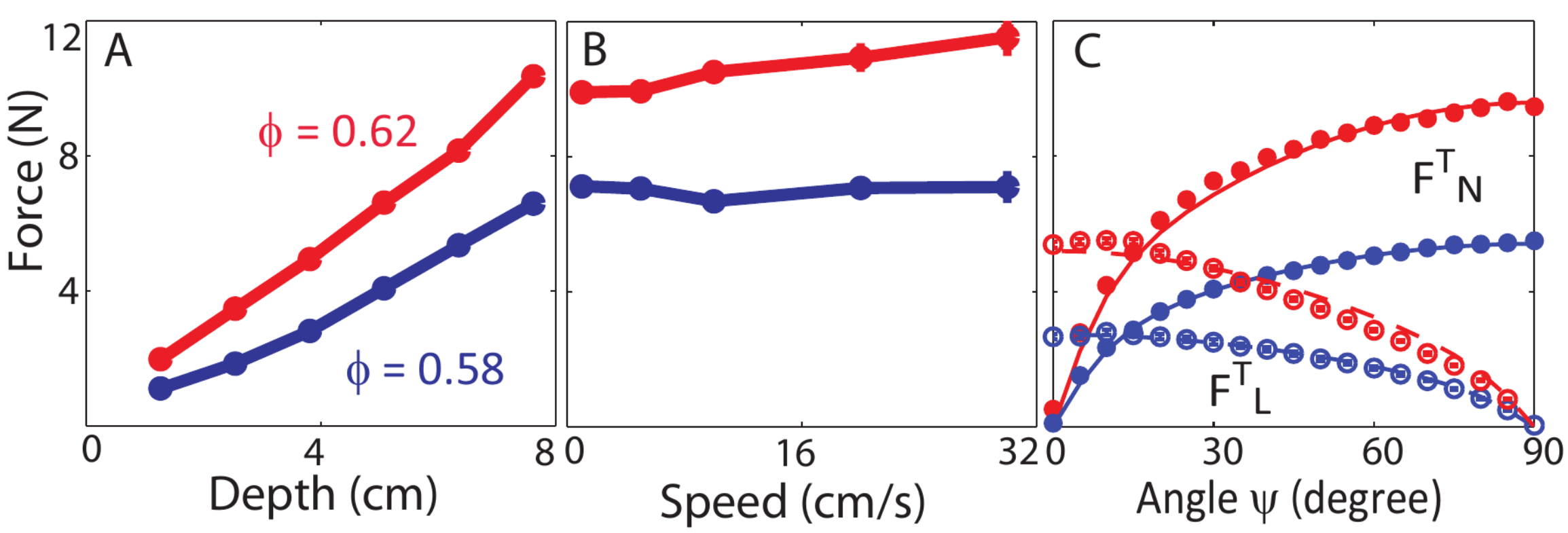

**Figure S2.** Force on a stainless steel cylinder as it is dragged at different depths, velocities, and angles through granular media of low and high φ . In (A) and (B) the cylinder is oriented perpendicular to the direction of motion. Average force vs. (A) depth and (B) speed for closely (red) and loosely (blue) packed preparations. (C) Average total forces (including sidewalls and end-faces) F T (closed symbols) and F T (open symbols) on the cylinder where the color of the N L symbols denote closely (red) and loosely (blue) packed preparations. Solid and dashed lines are model fits described by Equation S8. All forces are calculated by averaging instantaneous forces over steady state motion (e.g., see gray region in Fig. 3B); there is a 94% probability that

To characterize the relation between the measured values of normal and lateral drag force on the rod and its orientation $\psi$ relative to the direction of motion we developed an empirical fitting function inspired by soil mechanics (S5). Because there is a drag induced normal force perpendicular to the front half of the surface of the rod, there are two contributions to the resistive force in the horizontal plane of the motion. The first is the in-plane component of the normal force, characterized by $C_S$, and from the velocity independence of force (Fig. S2 B) we expect to be constant. The second is the in-plane component of the frictional force acting everywhere tangential to the surface of the rod characterized by $C_F$ . We expect the planar normal force to be larger than the frictional force because the friction between the grains and the rod is less than unity. The total force has contributions from the sidewall of the cylinder and the flat ends. For a cylinder with length $l$ and radius $r$, the total forces $F\ T$ and $F\ T$ can be $N\ L$ written as:

$$
\begin{aligned}
F_N^T &= 2lr(C_S \sin\beta_0 + C_F \sin\psi) + \pi r^2 C_F \sin(\psi) \\
F_L^T &= 2lrC_F \cos\psi + \pi r^2 C_S \sin\beta_0',
\end{aligned}
\tag{S8}
$$

where tan $\beta_0$ = cot $\gamma_0$ sin $\psi$ and tan $\beta_0'$ = cot $\gamma_0$ $\sin(\pi/2 - \psi)$ and $\gamma_0$ is a parameter related to the internal slip angle in granular media (S5).

The total forces $F_N^T$ and $F_L^T$ are determined as the averages over the shaded region in Fig. 3B. The fits to the measurements of $F_N^T$ and $F_L^T$ determine the three parameters $C_S$, $C_F$ and $\gamma_0$. These parameters are given in Table S1 and are used in Equation 1 to approximate the forces on the body $F_N$ and $F_L$ as the forces on the sidewalls of the cylinder without contributions from the flat ends (subtracting terms proportional to the flat end area, $\pi r^2$).

| Packing | $\phi$ | $C_S$ ($\mathrm{N/m^2} \times 10^{-4}$) | $C_F$ ($\mathrm{N/m^2} \times 10^{-4}$) | $\gamma_0$ (degree) |
|---|---|---|---|---|
| LP | 0.58 | 0.51 | 0.28 | 13.84 |
| CP | 0.62 | 0.77 | 0.59 | 12.21 |

Table S1: The coefficients of the rod drag model: quantities related to the yield stress of the granular media ($C_S$), the flow resistance coefficient ($C_F$), and the internal slip angle of the granular media under gravity ($\gamma_0$).

## Approximate Analytic Solution of the Resistive Force Model

To gain insight into how $\eta$ depends on kinematic parameters and features of the force model, we develop an approximate analytic expression assuming small amplitude oscillations such that $\theta$ is small, and small slip (high $\eta$) such that we approximate $\arctan(v_y/v_x)$ as $v_y/v_x$. Using Equation S2, $\psi$ can be approximated as

$$
\begin{aligned}
\psi &= \arctan(\frac{v_y}{v_x}) - \theta \\
\psi &\approx \frac{v_y}{v_x} - \theta \\
\psi &\approx \frac{1}{v_x}\frac{2A\pi v_w}{\lambda}\cos\frac{2\pi}{\lambda}(x + v_w t) - \frac{2A\pi}{\lambda}\cos\frac{2\pi}{\lambda}(x + v_w t),
\end{aligned}
\tag{S9}
$$

and since $\eta = v_x/v_w$,

$$
\psi = (\frac{1}{\eta} - 1)\frac{2A\pi}{\lambda}\cos\frac{2\pi}{\lambda}(x + v_w t). \tag{S10}
$$

The equation above shows that $\psi$ decreases as $\eta$ increases and $\psi$ oscillates along the body. For a sandfish swimming in a horizontal plane, the forces are only functions of $\psi$ and the sum of thrust and drag forces on the body is found from Equation S4.

Assuming large $\eta$, we approximate $P_N$ and $P_L$ as the low $\psi$ region of the curves in Fig. 3C and D. Thus the force per unit area is approximated as,

$$
\begin{aligned}
P_N &= C_N \psi \\
P_L &= C_L,
\end{aligned}
\tag{S11}
$$

where $C_N$ and $C_L$ are constants characterizing the granular medium's drag.

Using these approximations, the integral becomes:

$$
\begin{aligned}
F_{bx} &= \int_0^\lambda (C_N\psi\sin\theta - C_L\cos\theta)\sqrt{1+\tan^2\theta}2rdx \\
F_{bx} &= \int_0^\lambda \{C_N(\frac{1}{\eta}-1)\frac{2A\pi}{\lambda}\cos[\frac{2\pi}{\lambda}(x+v_wt)]\sin\theta - C_L\cos\theta\}\sqrt{1+\tan^2\theta}2rdx \\
F_{bx} &\approx \int_0^\lambda \{C_N(\frac{1}{\eta}-1)\frac{2A\pi}{\lambda}\cos[\frac{2\pi}{\lambda}(x+v_wt)]\frac{2A\pi}{\lambda}\cos[\frac{2\pi}{\lambda}(x+v_wt)] - C_L\}2rdx \\
F_{bx} &= [\frac{1}{2}C_N(\frac{1}{\eta}-1)(\frac{2A\pi}{\lambda})^2 - C_L]2r\lambda.
\end{aligned}
\quad \text{(S12)}
$$

Note that $v_wt$ does not appear in the final form because the integral over a wavelength is independent of the initial phase. If we assume that for small $\theta$ the head drag is a constant $F_{hx}$, and we set the net force to zero:

$$F_{xnet} = [\frac{1}{2}C_N(\frac{1}{\eta}-1)(\frac{2A\pi}{\lambda})^2 - C_L]2r\lambda - F_{hx} = 0. \quad \text{(S13)}$$

The wave efficiency $\eta$ can be found analytically. With $F_{hx} = 0$ the form of the solution is straightforward:

$$\eta = \frac{1}{1+\frac{2C_L}{C_N}(\frac{\lambda}{2\pi A})^2}. \quad \text{(S14)}$$

Estimating $C_N/C_L \approx 4$ and taking $A/\lambda \approx 0.2$ from the measured kinematics, Equation S14 gives $\eta = 0.76$. If the coefficient for head drag is estimated as $C_N$, then $\eta = 0.65$. The overestimate is likely due to the poor approximation of $\arctan(v_y/v_x) \approx v_y/v_x$, since for typical swimming speeds when $v_y$ is maximal, $v_y/v_x \approx 2.5$ and thus $\arctan(v_y/v_x) \approx 1.2$.

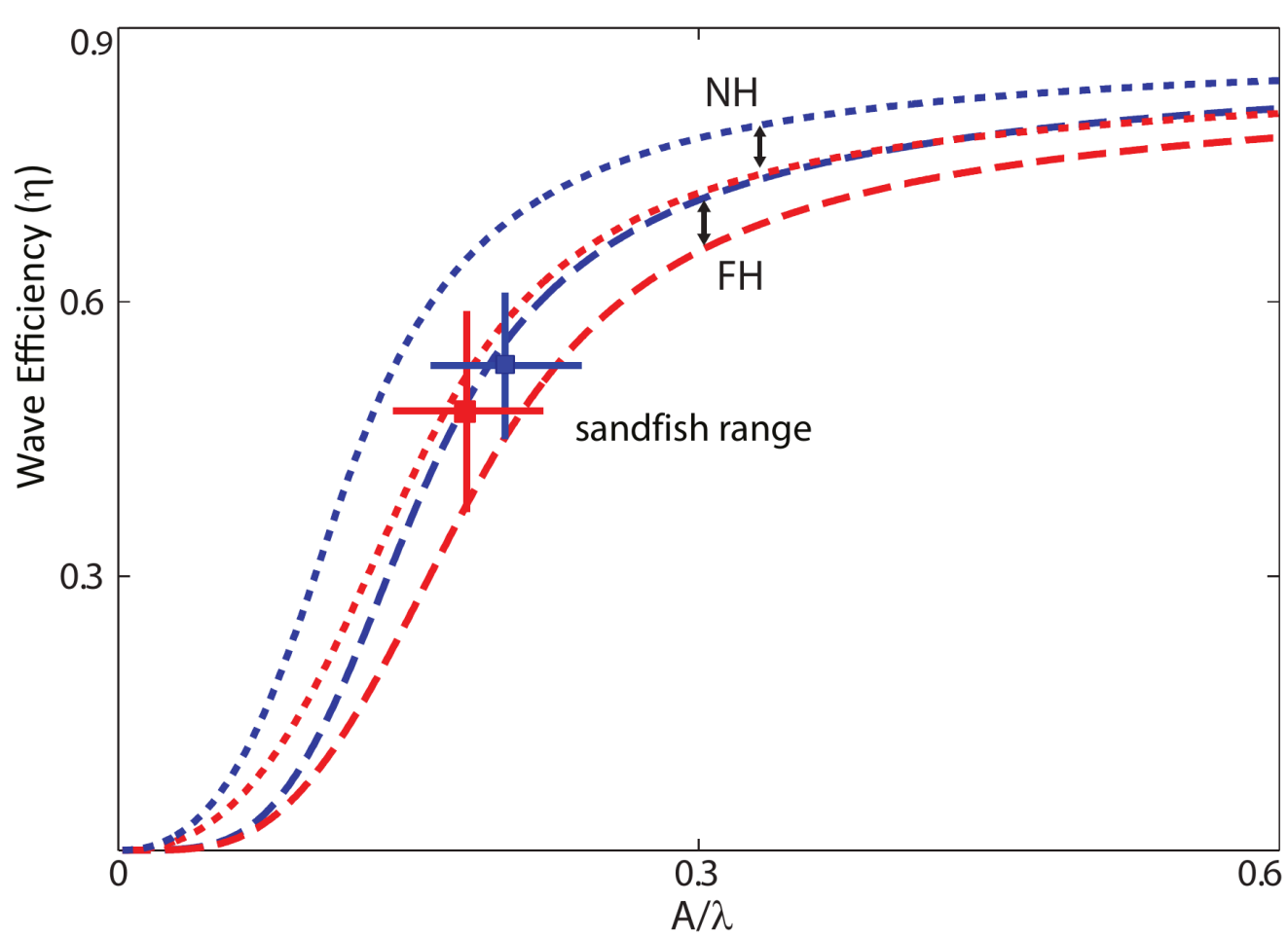

**Figure S3.** The wave efficiency predicted by the RFT model (numerical integration of the full model, Equation S7) is plotted against the spatial characteristic (A/λ) of the sinusoid. The blue and red curves correspond to predictions for LP and CP respectively. The red and blue squares with error bars show η measured from the kinematics of the sandfish in the x-ray experiments. Curves show model predictions using flat head (longer dash) and no head (shorter dash) limits.

## Assumptions and Error Analysis

There are several assumptions in our model of sandfish swimming, including those in the RFT, in the estimates of $F_N$ and $F_L$ from the rod drag experiments, in the empirical fitting function, in the body shape, and in the form of the drag on the head. 1) The RFT assumes that the organism swims in a horizontal plane, while in reality it moves into the material in a plane inclined at

$22°$ relative to the surface (S6). Since both drag and thrust are distributed over the body and are linear with depth (S4), we use the force on the central element as the average for the whole body. From x-ray data, the sandfish's head is $\sim$ 3 cm and its central element about $\sim$ 2 cm below the surface when the sandfish stops. Thus the force per area for the head is estimated to be 1.5 times larger than on the middle of the body. We assume $F_N(\psi)$ and $F_L(\psi)$ are the same as for horizontal drag. The projected force of gravity on the sandfish along the plane is neglected as it is small compared to the resistive force from the material (the ratio is estimated to be under 2%). We scaled the measured wavelength by $1/\cos(22°)$ = 1.08 to account for the projected wavelength on the horizontal plane of the x-ray images. 2) The forces used in the model are measured as the rod is dragged at constant velocity. In reality the velocity of each point on the body varies sinusoidally as the animal moves forward. 3) The diameter of the sandfish changes from head to tail, while a single diameter of 1.75 cm is used for the model. We approximated the sandfish as a single cylinder of fixed radius for simplicity of the model and calculation (i.e. time invariance of Equation S4). Since in the animal the radius shrinks near the head and tail, the surface of the body is not parallel to the spine of the sandfish. Our model predicts that the change in the diameter near the head (which results in a smaller $\psi$ than in the uniform cylinder), causes decreased $F_N$ and increased $F_L$. The taper of the tail results in larger $\psi$ compared to the uniform cylinder which results in increased $F_N$ and decreased $F_L$. We hypothesize that these effects result in a cancelation that makes the uniform cylinder a good approximation. 4) Since we did not have an estimate for the drag on the shovel shaped head of the sandfish, we computed $\eta$ for two cases of head drag. We considered drag on the head zero (NH in Fig. 4A and S3) or that of a flat disk of radius comparable to the sandfish (FH). These predictions in conjunction with the model for body drag bound the experimentally measured $\eta$. With the model assumptions in place, the major contribution to the uncertainty in $\eta$ comes from the experimentally measured run to run variation in the spatial parameters $A$ and $\lambda$.

## Supplementary References